\documentclass[epjST]{svjour}
\usepackage{graphicx}
\usepackage{url}
\usepackage{hyperref}
\begin{document}
\title{Monetary economics from econophysics perspective}
\author{Victor M. Yakovenko\thanks{\email{http://physics.umd.edu/\~{}yakovenk/}}}
\institute{Department of Physics, CMTC and JQI, University of Maryland, College Park, Maryland 20742-4111, USA}
\abstract{
This is an invited article for the \emph{Discussion and Debate} special issue of The European Physical Journal Special Topics on the subject ``Can Economics Be a Physical Science?''  The first part of the paper traces the personal path of the author from theoretical physics to economics.  It briefly summarizes applications of statistical physics to monetary transactions in an ensemble of economic agents.  It shows how a highly unequal probability distribution of money emerges due to irreversible increase of entropy in the system.  The second part examines deep conceptual and controversial issues and fallacies in monetary economics from econophysics perspective.  These issues include the nature of money, conservation (or not) of money, distinctions between money vs.\ wealth and money vs.\ debt, creation of money by the state and debt by the banks, the origins of monetary crises and capitalist profit.  Presentation uses plain language understandable to laypeople and may be of interest to both specialists and general public.
}

\maketitle

\section{A personal path from physics to economics}  \label{Sec:Personal}

First I would like to explain the style and purpose of this article.  I have published a number of research and review papers on econophysics \cite{Yakovenko-Web} and have given many talks and colloquia about it.  These talks are usually followed by informal discussions with the attendees, which cover broader material  than in an academic presentation.  People often ask me whether I published such a broader discussion somewhere, and the answer has been negative until now.  I would like to use the occasion of this \emph{Discussion and Debate} special issue to describe my personal research motivation and path to melding physics and economics.  I would also try to articulate and clear up some confusion related to deep conceptual issues in monetary economics from the standpoint of a theoretical physicist.  The style of this paper is quite informal, and the primary focus is on conceptual foundation, rather than technical details of mathematical models and empirical data.  

In 1978 I became an undergraduate student at Moscow Institute of Physics and Technology (MIPT).  I wanted to learn theoretical physics, so I started reading the famous 10-volume \emph{Course of Theoretical Physics} by L.~D.~Landau and E.~M.~Lifshitz.  In due time I advanced to volume 5 \emph{Statistical Physics} \cite{Landau-5}.  Statistical Physics is quite different from other branches of physics, such as Mechanics and Electrodynamics, where time evolution of particles and fields is deterministically defined by equations of motion starting from initial conditions.  Even in Quantum Mechanics, where measurements involve probabilities, time evolution of the wave function is governed by the deterministic and unitary equation of motion.  In contrast, Statistical Physics studies big ensembles of objects (such as atoms and molecules) and describes them using probability theory.  

At the very beginning, in Ch.~1.4 ``The Role of Energy'', the book \cite{Landau-5} draws attention to those random variables that obey conservation laws.  The prime example of such a variable is energy $E$.  If we divide the system into two subsystems, the total energy $E$ is the sum of two parts: $E=E_1+E_2$.  On the other hand, the probability distribution $P(E)$ of energy $E$ is the product of the two probability distributions: $P(E)=P(E_1)\,P(E_2)$.  In other words, energy is additive, whereas probability is multiplicative.  The general solution of these two equations is the exponential function $P(E)\propto e^{-E/T}$, where $T$ is a coefficient called temperature.  (The Boltzmann constant is set to unity $k_B=1$, so $T$ has the dimensionality of $E$.)  It is known as the Boltzmann-Gibbs distribution of energy in physics.

I was stunned that such a general result is derived seemingly from nothing, just using the basics of probability theory.  So, I thought this mathematical derivation should be applicable to a much broader class of systems, which have nothing to do with physics, as long as they are statistical ensembles and have a conserved variable.  As a prime example, I thought of the economy, which is surely a big statistical ensemble with millions of interacting agents.  But is there a conserved variable?  Goods and services, the main purpose of the economy, are clearly not conserved, because they are produced and consumed.  But in a monetary economy, goods and services are exchanged for money.  Ordinary economic agents can only receive and give money to other agents, but are not permitted to ``manufacture'' money.  (This would be criminal counterfeiting.)  In other words, the agents can grow apples on apple trees (so apples are not conserved), but cannot grow money on ``money trees''.  Suppose an agent 1 with the initial money balance $m_1$ transfers money $\Delta$ to agent 2 with the balance $m_2$ and receives apples in exchange.  Agent 1 eats the apples and they disappear, but the new money balances of the agents now are $m_1'=m_1-\Delta$ and $m_2'=m_2+\Delta$.  Clearly, money satisfies a local conservation law $m_1'+m_2'=m_1+m_2$ in transactions among agents.  While conservation laws in physics follow from fundamental space-time symmetries, conservation of money is the law of accounting.

So I concluded, by analogy with the derivation in \cite{Landau-5}, that the probability distribution $P(m)$ of money $m$ among the agents of an economic system in statistical equilibrium should be given by the exponential function $P(m)\propto e^{-m/T_m}$, where $T_m$ is the money temperature.  The latter is equal to the average amount of money per agent: $T_m=\langle m\rangle=\int_0^\infty dm\,m\,P(m)/\int_0^\infty dm\,P(m)=M/N$, where $M$ is total money and $N$ is the number of agents.  The exponential probability distribution $P(m)$ implies that there are few rich agents with high money balance and many poor agents with low balance, so the distribution is highly unequal, even if the agents are statistically equal.  This is a shocking conclusion for social sciences, where inequality is usually attributed to intrinsic differences between agents, but in a gas it results from probabilistic collisions between identical molecules.  Although I came to these conclusions as a physics undergraduate, I waited for 20 years to get tenure before publishing this idea in \cite{Yakovenko-2000} with my student Adrian Dr\u{a}gulescu, because I was not sure how my physics colleagues would react to such a radical proposition.

Besides, I wanted to finish reading the remaining volumes of the \emph{Course of Theoretical Physics}.  Relevantly, volume 10 \emph{Physical Kinetics} \cite{Landau-10} introduced me to the Boltzmann kinetic equation, which describes evolution of $P(m)$ in time $t$,
\begin{eqnarray}
  &&\frac{dP(m)}{dt}=\int\!\!\!\!\int\{
    -f_{[m,m']\to[m-\Delta,m'+\Delta]}P(m)P(m')
\label{Boltzmann}  \\
  &&+f_{[m-\Delta,m'+\Delta]\to[m,m']}
  P(m-\Delta)P(m'+\Delta)\}\,dm'\,d\Delta.
\nonumber
\end{eqnarray}
Here $f_{[m,m']\to[m-\Delta,m'+\Delta]}$ is the probability of transferring money $\Delta$ from an agent with money $m$ to an agent with money $m'$ per unit time.  This probability, multiplied by the
occupation numbers $P(m)$ and $P(m')$, gives the rate of transitions from the state $[m,m']$ to the state $[m-\Delta,m'+\Delta]$.  The first term in Eq.\ (\ref{Boltzmann}) gives the depopulation rate of the state $m$.  The second term in Eq.\ (\ref{Boltzmann}) describes the reversed process, where the occupation number $P(m)$ increases.  When the two terms are equal, the rates of direct and reversed transitions are equal, so the probability distribution is stationary: $dP(m)/dt=0$.  This is called the principle of detailed balance.

In physics, the fundamental microscopic equations of motion obey the time-reversal symmetry, so the probabilities of direct and reversed processes are exactly equal:
\begin{eqnarray}
  f_{[m,m']\to[m-\Delta,m'+\Delta]}=f_{[m-\Delta,m'+\Delta]\to[m,m']}.
\label{reversal}
\end{eqnarray}
When Eq.\ (\ref{reversal}) is satisfied, the detailed balance condition for Eq.\ (\ref{Boltzmann}) reduces to the equation $P(m)P(m')=P(m-\Delta)P(m'+\Delta)$, because the factors $f$ cancels out.  The general solution of this equation is again the exponential function $P(m)\propto e^{-m/T_m}$, i.e.\ the Boltzmann-Gibbs distribution is the stationary solution of the dynamical Boltzmann kinetic equation (\ref{Boltzmann}).  Notice that, while the transition probabilities (\ref{reversal}) are determined by the dynamical rules of the model, the equilibrium Boltzmann-Gibbs distribution does not depend on the dynamical rules at all.  Amazingly, it is possible to find the stationary distribution without knowing details of the dynamical rules (which are not so clear in economics), as long as the symmetry condition (\ref{reversal}) is satisfied.

This brings up the question whether dynamical rules for economic transactions have the time-reversal symmetry.  In physics, time-reversal invariance is a fundamental symmetry, but, in economics, it is not necessarily the case.  As an example of a time-reversible operation, suppose I pay 5 dollars to buy apples and then receive 5 dollars for selling oranges from my farm, so my balance returns to the initial value.  These two transactions correspond to the left and right sides of Eq.~(\ref{reversal}) with $\Delta=5$.  Moreover, if $f(\Delta)$ in Eq.~(\ref{reversal}) depends only on $\Delta$, but not on $m$ and $m'$ (or depends only on $m+m'$), then the probabilities of money flow in both directions are equal, and time-reversal symmetry is satisfied.  As an example of a non-time-reversible operation, suppose I pay 5\% of my money balance, i.e.\ $\Delta=0.05\,m$, to buy apples, and then another agent pays me 5\% of her money balance, i.e.\ $\Delta'=0.05\,m'$, for oranges from my farm.  Obviously, my balance does not return to the initial value, and the left and right terms in Eq.~(\ref{reversal}) are not equal.  The first process is additive, because $\Delta$ does not depend on $m$, whereas the second process is multiplicative, because $\Delta$ is a percentage of $m$.  Models of the second kind were introduced in econophysics literature in \cite{Redner-1998,Chakraborti-2000} and earlier in sociology \cite{Angle-1986}.  The probability distribution of money $P(m)$ in these models was found to be Gamma-like, i.e.\ not exponential, although its high-end tail is still exponential.  So, despite the elegant argument presented in \cite{Landau-5}, conservation of energy or money is not, by itself, sufficient to produce the exponential distribution, but microscopic time-reversal symmetry is sufficient, as explained in \cite{Landau-10}.  Whether additive, multiplicative, or mixed models are appropriate for economics is subject to debate, not to be pursued here, but discussed in some detail in reviews \cite{Yakovenko-2009,Chakrabarti-2013}.

If the transaction amounts $\Delta$ are small, the integral Boltzmann equation (\ref{Boltzmann}) for the probability distribution $P(m,t)$ can be reduced to the partial differential Fokker-Planck equation \cite{Landau-10}, also known as the Kolmogorov forward equation,
\begin{equation}
   \frac{\partial P}{\partial t}=\frac{\partial}{\partial m} \left[AP
   + \frac{\partial(BP)}{\partial m}\right], 
   \quad A=-{\langle\Delta\rangle \over dt}, 
   \quad B={\langle\Delta^2\rangle \over 2 dt}.
\label{diffusion}
\end{equation}
The coefficients $A$ and $B$, known as the drift and diffusion, are the first and second moments of money balance changes $\Delta$ per time increment $dt$.  Notice that Eq.~(\ref{diffusion}) is linear with respect to $P(m,t)$, whereas Eq.~(\ref{Boltzmann}) is nonlinear.  A stationary solution $\partial_tP=0$ of Eq.\ (\ref{diffusion}) is obtained by setting the probability flux, which is the expression in the square brackets, to zero.  For an additive process, the coefficients $A$ and $B$ are constants independent of $m$, so the stationary distribution is again exponential $P(m)\propto e^{-m/T_m}$.  Here the money temperature $T_m=B/A$ is expressed in terms of $B$ and $A$ similarly to the Einstein relation between temperature, diffusion, and mobility.  A well-known example in physics is the barometric distribution of gas density $P(z)\propto e^{-\mu gz/T}$ versus height $z$ in the presence of gravity, where $\mu$ is the mass of a molecule, $g$ is the gravitational acceleration, and $\mu gz$ is the gravitational energy of a molecule.  The competition between downward drift due to gravity and spreading around due to diffusion results in the stationary exponential distribution $P(z)$.  It represents statistical equilibrium as opposed to mechanical one, where two forces acting on an object balance each other.  Much of the current thinking in economics revolves around mechanical equilibrium between supply and demand, but the concept of statistical equilibrium is largely unfamiliar to most economists.

Besides tenure, another important factor that nudged me to proceed with the paper \cite{Yakovenko-2000} was invention of the term \textit{econophysics} by Eugene Stanley \cite{Stanley-1996} in a discussion with Bikas Chakrabarti at the conference \textit{Dynamics of Complex Systems} in Kolkata in 1995 \cite{Chakrabarti-history}.  Although some physicists dabbled in economics before (see review \cite{Yakovenko-2009}), the new term \textit{econophysics} prompted consolidation and explosive growth of research by physicists this area.  However, the road was very bumpy at the beginning.  

The paper \cite{Yakovenko-2000} was originally submitted to Physical Review Letters.  It was rejected based on a negative referee report stating that it is known since the days of Vilfredo Pareto that distributions in economics are power laws and are not exponential.  Interestingly, the argument was empirical, rather than theoretical, so we decided to re-examine empirical data.  We did not succeed in finding data on distribution of money balances, but found plenty of data on income distribution from several government agencies.  When we plotted the data from the U.S.\ Census Bureau in log-linear scale, the data points fell on a straight line, confirming the exponential law, which we reported in \cite{Yakovenko-2001a}.  Then, using the data from the U.S.\ tax agency, the Internal Revenue Service (IRS), extending to much higher incomes that the Census data, we found that the high end of the distribution does follow the Pareto power law, whereas the majority of population with lower incomes is described by the exponential law \cite{Yakovenko-2001b}.  Thus, the whole income distribution can be decomposed into the lower class with an exponential distribution and the upper class with a power law, indicating the two-class structure of the society \cite{Yakovenko-2003}.  We confirmed these results by an extensive analysis of income data for about 20 years from the IRS \cite{Yakovenko-2005a}.  This paper was also originally submitted to Physical Review Letters and received generally positive referee reports, but was still rejected on the grounds that ``it is not physics''.  Nevertheless, four years later, Reviews of Modern Physics accepted the paper \cite{Yakovenko-2009}, albeit not without a fight with referees.  In this paper and in \cite{Banerjee-2010}, we proposed a unified description of the power-law and exponential classes in income distribution based on the Fokker-Planck equation (\ref{diffusion}) with co-existing additive and multiplicative stochastic processes.

Reaction to these studies from mainstream economists was not enthusiastic, to put it mildly.  At a conference lunch in Aix-en-Provence in 2002, a distinguished professor of economics from Princeton University told me: ``Inequality is not something we are interested in economics''!  But after the big financial crisis of 2008, the existence of the two social classes, 99\% vs.\ 1\%, became widely recognized, and debate on inequality took central stage, culminating in the rock-star reception of Piketty's book \cite{Piketty}.  In 2014, the Science magazine dedicated a special issue to \textit{The Science of Inequality}, where the physics-inspired theory of inequality was covered in the article \cite{Cho}.

In physics, a very important concept in dealing with statistical distributions is entropy.  Ludwig Boltzmann proved in the 19th century that the exponential distribution of energy in statistical equilibrium can be derived from the principle of maximal entropy, subject to the constraint of a fixed total energy in the system.  According to Boltzmann, the entropy $S=\ln W$ is the logarithm of the multiplicity $W$, which is the number of combinatorial ways for distributing the total energy $E$ (or money $M$) among the molecules in a gas (or agents in the economy).  For example, there is only one way to divide $E$ or $M$ equally among $N$ molecules or agents: $m_i=M/N$, so $W=1$ and $S=0$.  In contrast, in unequal distribution, one agent may have a balance $m_1=5$ and another agent $m_2=10$, or, conversely, their balances may be $m_1=10$ and $m_2=5$, so the same overall distribution can be obtained in several different ways.  Thus, $W>1$ and $S>0$ for unequal distribution, and the maximal values of $W$ and $S$ are achieved for the exponential distribution.  It follows from the kinetic equation (\ref{Boltzmann}), as proven by Boltzmann, that the entropy $S$ increases monotonously in time until it reaches the maximal value \cite{Landau-10}.  So, the exponential distribution of money is a consequence of the maximal entropy principle \cite{Yakovenko-2013}, which is the basis of the second law of thermodynamics.  This principle can be also applied to global inequality, as measured by energy consumption per capita in different countries \cite{Banerjee-2010}.  Analysis \cite{Lawrence-2013} of the data from the U.S.\ Energy Information Agency (EIA) shows that, in the last 30 years, global energy inequality has been decreasing from a very unequal distribution between developed and developing countries toward the more equal exponential distribution, in agreement with the maximal entropy principle.  In this approach, entropy saturation at the exponential distribution is responsible for the current slow-down of the global economic activity, manifestations of which are recognized as the ``global economic ice age'' \cite{Ice-Age} or ``secular stagnation'' \cite{Larry-Summers}.  The mechanism is similar to the ``thermal death of the Universe'' widely discussed in physics in the 19th century.

\section{Debate about money}  \label{Sec:Money}

Having discussed statistical distribution and inequality in an economic ensemble, now I would like to focus on the conservation law of money, which initially motivated this study.  Generally, conservation laws play very important role in physics, e.g.\ by imposing constraints that ensure stability of a system, but economics literature pays less attention to them.  Every time I give a talk and describe conservation of money in transactions among agents, somebody exclaims: ``But money is not conserved!''  Then I ask how money is not conserved and receive all kinds of answers.  So, in the second part of the paper, I examine various mechanisms for putative non-conservation of money, after a discussion of the nature of money.  As a reference, I use three Chapters ``Money and Banking'', ``How Banks Create Money'', and ``The Federal Reserve Banks and Monetary Policy'' from the fairly standard undergraduate economics textbook \cite{McConnell}.  This discussion also addresses confusion in econophysics literature between the concepts of money and wealth.

\subsection{The nature of money, conservation law, and boundary conditions}  \label{Sec:Nature}

First, one may ask why do we need to use money.  After all, the ultimate purpose of the economy is production of goods and services.  In principle, barter can be used for a direct exchange of, say, apples for oranges between two agents.  However, barter is increasingly impractical for a cyclic exchange, say, of apples, oranges, and cherries among three agents and for higher-order exchanges.  Thus, a natural idea is to introduce digital tokens, called money, so that agent 1 receives money (rather than physical products) in exchange for delivering apples to agent 2.  Some time later, agent 1 gives money to agent 3 and receives oranges in exchange.  The net result is that agent 1 exchanged apples for oranges, but not necessarily with the same agent, as in barter.  This makes the economy much more flexible than in barter, where flow of goods must be balanced on pairwise basis.  Now an agent only needs to balance a contribution of apples to the society as a whole, for which the agent is paid money, with the benefit of receiving oranges from the society, for which the agent pays money. 

This consideration shows that \emph{money is a digital accounting tool needed to prevent free riding}.  Money is a digital token (meaning it is expressed as a number) indicating how much an agent contributed to the society, which entitles the agent to receive a commensurate benefit from the society.  If such a record is not maintained, agents would engage in free riding, i.e.\ they would receive benefits from the society (from the other agents) without contributing to the society in return, which would quickly ruin economic relations.  Money is not needed for pairwise exchanges of goods in barter, but a universally accepted digital tool, i.e.\ money, is necessary to enable many-body exchanges of goods in the ensemble of agents who don't even know each other.

Since the purpose of money is to prevent free riding, \emph{money must necessarily be conserved}.  It means that agents should be not be able to manufacture (or destroy) the digital money tokens on their own (unlike apples), but should only receive them from and give them to other agents.  Otherwise, the agents would manufacture the digital money tokens and use them to receive benefits from the society without contributing goods and services in return, which is free riding.  So, conservation of money is not a peculiar artifact, but the fundamental principle of accounting on which money functionality is based.  Curiously, conservation of money is not mentioned explicitly in economic textbooks \cite{McConnell}, although the three commonly stated properties of money: medium of exchange, measure of value, and store of value, only make sense only if money is conserved.  Indeed, medium of exchange means that money is exchanged for goods in transaction among agents, the value of goods is measured by the money paid for them, and storing money entitles to receive future benefits.  These feature would become meaningless, if the agents were able to manufacture money on their own.

But if the agents cannot manufacture money, how would money be brought into the system?  Instead of going through the history of various rare and scarce objects (notably gold) serving as money, let us focus on the present-day fiat money, which are, essentially, digital accounts declared to be money.  As an example, let us discuss a Local Exchange Trading System (LETS) \cite{LETS}, as a bottom-up way to create a monetary system.  LETS participants start with zero initial balances $\tilde m_i=0$.  When agent 1 provides a service to agent 2, their balances increase and decrease by the same amount: $\tilde m_1'=+\Delta$ and $\tilde m_2'=-\Delta$.  So, the conservation law is satisfied in the algebraic sense: $\tilde m_1+\tilde m_2=0=\tilde m_1'+\tilde m_2'$.  As the agents continue trading, their balances grow in positive and negative directions starting from zero so that $\langle\tilde m\rangle=0$, and it looks like this system does not require initial creation of money.  

However, the conservation law by itself does not prevent free riding.  Some agents $i$ can accumulate unlimited negative balances $\tilde m_i<0$ by receiving services and not contributing in return.  This can be prevented by imposing a boundary condition on the negative side, i.e.\ by demanding that $\tilde m_{\rm min}<\tilde m_i$, where $\tilde m_{\rm min}<0$.  An agent whose balance reached $\tilde m_{\rm min}$ is not permitted to receive further services until contributing services and, thus, increasing the balance above $\tilde m_{\rm min}$.  One may ask whether a similar boundary condition should be imposed on positive balances: $\tilde m_i<\tilde m_{\rm max}$ with $\tilde m_{\rm max}>0$.  If the boundary conditions are symmetric $\tilde m_{\rm min}=-\tilde m_{\rm max}$, then it is obvious that the distribution of money $P(\tilde m)$ would be symmetric around zero: $P(\tilde m)=P(-\tilde m)$.

However, one may argue against imposing the upper boundary condition $\tilde m_{\rm max}$.  Why stop agents from contributing more to the society, which results in the increase of their positive money balances?  An agent can be proud of a high positive balance, which confirms  great contribution to the society (the opposite of free riding).  So, suppose the upper limit is not imposed, i.e.\ $\tilde m_{\rm max}=\infty$.  Now the boundary conditions are asymmetric, and, obviously, $P(\tilde m)$ would be also asymmetric.  At this point, LETS coordinators may realize that this asymmetric system can be simplified by shifting all money balances to introduce new money $m_i=\tilde m_i-\tilde m_{\rm min}$.  It is equivalent to giving each agent the positive amount $-\tilde m_{\rm min}$.  Now the boundary condition shifts to zero, so the new money must be positive $m_i>0$ with the mean $\langle m\rangle=|\tilde m_{\rm min}|=M/N=T_m$, which we called the money temperature in Sec.\ \ref{Sec:Personal}.  LETS coordinators may also realize that the new money $m_i$ could have been introduced from the beginning by allocating the equal initial balances $m_i=M/N$ to all agents and imposing the boundary condition $m_i>0$ to prevent free riding.  Since money $m$ is positive, it can also implemented using symbolic physical objects such as gold, but it would have been difficult to implement ``negative gold'' for $\tilde m$.  In practice, positive fiat money $m$ on the national scale is usually created in a top-down manner by the central monetary authority of the state, as discussed in Sec.\ \ref{Sec:State} and emphasized by the Modern Monetary Theory (MMT) \cite{MMT,MMT-family,Wray}.

If the agents are given the equal initial money balances, would their balances stay approximately equal, once they start trading?  One might naively expect that the answer is positive, given that all agents are treated equally in a statistical sense.  However, agent-based computer simulations \cite{animation} show that the answer is actually negative.  In a simple model where $\Delta=1$ for each transaction \cite{animation}, the initial delta-function distribution $P_0(m)=\delta_{m,\langle m\rangle}$ starts broadening in a Gaussian manner, then becomes asymmetric because of probability accumulation at the boundary $m=0$, and finally attains the stationary exponential form $P(m)\propto e^{-m/T_m}$.  This example teaches several lessons.  First, the initial state of perfect equality is totally unstable once trading starts.  As explained in Sec.\ \ref{Sec:Personal}, the state of perfect equality has zero entropy, so the system spontaneously evolves toward unequal state with higher entropy.  Second, the lower boundary condition $m>0$ is absolutely crucial for achieving a stationary probability distribution $P(m)$.  If the boundary condition $m>0$ were not imposed (i.e.\ negative money balances and free riding were permitted), then the system would never achieve any equilibrium, and $P(m)$ would never stabilize.  Some agents would keep accumulating negative money balances for receiving services from other agents, who would keep accumulating positive money balances.  This situation looks similar to the present-day imbalances in international trade, e.g.\ between USA and China, Germany and the rest of Europe, etc.  The \emph{laissez-faire} school of economics argues that, if all restrictions and regulations are removed, the economy would come to a natural equilibrium and should be left there.  However, this argument is misleading because \emph{an equilibrium cannot be achieved without imposing boundary conditions, i.e.\ regulations}.

The situation where the agents are given the equal initial money balances may look like an artificial exercise, but it actually happened in history.  A senior physics colleague from Germany told me that, after the World War II, when German economy was completely destroyed, the Allied Powers allocated an equal amount of new money to each German citizen in order to restart the economy.  Needless to say that the distribution of money in Germany now is not equal, so the inequality developed spontaneously out of the initially equal state.  Another example is the Russian revolution of 1917, which divided land equally among peasants.  However, forced confiscation of agricultural products by the Bolshevik government led to famine, so in 1921 Vladimir Lenin introduced the New Economic Policy (NEP) \cite{NEP} permitting limited market relations.  As a result, inequality started to develop among peasants via land acquisition by richer \emph{kulaki} from poorer \emph{batraki}.  In this case, inequality developed by redistribution of a positive conserved physical resource (land) rather than conserved digital variable (money), but mathematically these problems are the same.  Joseph Stalin terminated NEP in 1928 and replaced individual land ownership with collective farms.

\subsection{Distinction between physical and monetary layers of the economy}  \label{Sec:Layers}

The flow of goods and money in the monetary economy can be decomposed into two circuits or layers.  The physical layer includes production, transfer, and consumption of physical objects, goods, and services.  This layer is ultimately governed by physical laws and constraints, including energy, natural resources, environment, etc.  Objects at the physical layer (goods and services) are generally not conserved.  In contrast, the monetary layer represents flow of money among the agents in exchange for goods and services.  As argued in Sec.\ \ref{Sec:Nature}, money is an accounting tool represented by pure numbers, which are nothing but digital bits of information.  Thus, the monetary layer is the informational layer of the economy, in contrast to the physical layer.  Any information must have some physical carrier (e.g.\ paper banknotes, gold coins, or magnetic records on computer hard drives), but physical nature of the carrier is irrelevant, and information is subject to its own laws (e.g.\ the laws of accounting).  

The monetary layer also conveys information to the agents, guiding them toward optimal (in an ideal case) allocation of labor and resources.  Many people want to be artists and musicians, but society needs plumbers and garbage collectors a lot more to function effectively.  So, ideally, the agents would find more money earning opportunities in the areas of dire societal need.  The no-free-riding constraint of money imposes discipline and forces the agents to do what society needs, not necessarily what they like.

The two layers are coupled, because physical goods and digital money flow in the opposite directions in transactions between agents.  But it is very important to realize that the objects belonging to different layers cannot be converted into each other.  An (imperfect) analogy in physics is with fermions (e.g.\ electrons) and bosons (e.g.\ photons), which interact, but cannot be converted into each other.  A failure to distinguish between the two fundamentally different layers is pervasive in economics and among general public.  It originates from misinterpretation of what happens when an agent spends digital money to acquire a physical product.  From an individual point of view, money disappears and the product appears, so it looks like money is transformed into the product.  However, this is actually a transaction between two agents, so money does not disappear, but is transferred to another agent, so it is conserved, and the product arrives from that agent in counterflow of money and goods.  The failure to make this simple observation leads to various fallacies.

\subsubsection*{Fallacy \#1: Money grows as a result of investment}  

Once the chair of the economics department of the University of California at Santa Barbara came to my colloquium at the Kavli Institute for Theoretical Physics.  He objected to conservation of money, so I asked for an example (the audio-video record of this conversation is available online \cite{KITP}).  He proposed that I invest my money: spend \$100 to buy gold, then sell gold a year later for \$200, and, \emph{voila}, my money has increased, so it is not conserved.  My response was that it is necessary to consider all parties involved in these transactions.  Besides me, agent 1, there was also agent 2 who gave me gold and took my \$100.  Then, there was agent 3 who gave me \$200 in exchange for gold.  Overall, it was a three-body transfer of money mediated by counterflow of gold at the physical layer, and the sum of money balances of all three agents did not change, $m_1+m_2+m_3={\rm const}=\$300$. So, the conservation law of money is satisfied in investment.

The same fallacy is perpetuated on a much grander scale in pension funds investments affecting millions of people.  When I became a faculty member at the University of Maryland in 1993, I signed up for a 401(k) pension fund.  A representative recommended me to buy stocks now, so that I would receive a lot more money by selling them later, when I am ready to retire.  I asked where this extra money would come from to pay the higher price for my stocks.  He did not have better answers than ``everybody invests in stock market, because it is such a great thing'' and ``historically, stocks outperform everything else''.  (Later I heard similar mantra ``prices of houses never decrease''.)  These arguments did not satisfy my analytical mind, so he recommended me to read the book \cite{Dent-1993}.  The arguments in the book based on demographics did make sense.  The demographic tsunami wave of baby boomers progressively reaching the age of 45 and massively ``saving'' money for retirement by investing in stocks would create a stock market bubble starting from 1995.  It was predicted in the book published in 1993, as reflected by its title \cite{Dent-1993}.  Some 15 years later, reaching the age around 60, the baby boomers would start to retire and sell their investments, causing the market to crash.  On page 16, the book says ``the next great depression will be from 2008 to 2023''.  This is a stunning prediction of the 2008 financial crisis published 15 years in advance.  So, the baby boomers would be buying stocks during the bubble of late 1990s at high prices and selling them after the market crash of 2008 at low prices, which is the opposite of good investment.  Only the massive infusion of several trillion dollars by the Federal Reserve stopped financial system and stock market from collapse after 2008, but it remains to be seen what will happen next.  But the answer to my question seems to be that the money, hopefully, would come from the Fed.

\subsubsection*{Fallacy \#2: Money grows as a result of production}  

When people hear about conservation of money, they often respond along the following lines.  We work very hard to make our lives better, so our collective wealth increases in time as a result of production, and, since money is some sort of measure of wealth, money should increase too.  This is a typical failure to distinguish between physical and monetary layers.  Undoubtedly, physical capacity, as measured by the number of cars, homes, computers, etc., increases as a result of production.  However, this expansion of the physical layer does not translate directly into expansion of the monetary layer, because these layers have different nature and follow different rules.  For example, my first personal computer purchased in late 1980s had disk capacity of 20 MegaBytes.  Today, a typical computer can easily have disk capacity of 2 TeraBytes.  So, the disk storage capacity has increased by the factor of $10^5$.  Yet, today's computer is cheaper than my first computer, even without correction for inflation.  So, the humongous increase of physical storage capacity did not increase monetary value of a computer at all.  Instead, the price of storage per MegaByte has decreased by the factor of $10^{-5}$, and the annual dollar revenue of magnetic recording industry has been flat for 30 years.  This kind of argument is often brought up by the monetary school of economics \cite{McConnell}.

Another example of this fallacy can be seen in the debate about Greek debt.  It is often said that productivity in Greece is too low, and, if only Greece managed to increase production, then it would have surely been able to pay its debt.  Suppose Greece specializes in growing oranges and, by working very hard, doubles production, i.e.\ expands its physical layer.  The problem, however, is that the creditors of Greece do not want the debt to be paid in oranges; they want it to be paid in monetary units, dollars or euros.  So, Greece would need to find other parties who have enough money and are willing to pay this money for oranges, and that is a problem.  Any salesman can tell that sale of a product is often more difficult than production.

\subsubsection*{Fallacy \#3: Economic models without money}

Let us explore the analogy between two layers of the economy and fermions/bosons in physics a bit further.  In physics, we can ``integrate out'' photons, i.e.\ eliminate them as mediators, and obtain direct Coulomb interaction between electron.  Alternatively, we can ``integrate out'' electrons and describe an electronic material as an effective medium for propagation of photons.  

In economics, we can ``integrate out'' money, i.e.\ eliminate them as intermediate steps, and reduce an economic model to direct exchanges of physical products described by Leontief-style input-output matrices.  This approach is quite popular in economics, but there is a problem with such models without money.  Most economic crises in recent memory were crises of the monetary layer, not the physical layer.  If money is eliminated from a model, under the assumption of ideal functioning of monetary system, then, as a matter of principe, such a model cannot predict, describe, and explain monetary crises of the economy.

Alternatively, we can ``integrate out'' products and reduce a model to money transfers among agents.  This is the approach described in Sec.\ \ref{Sec:Personal}.  Even though such model only keeps track of money movements, it is assumed that there is corresponding counterflow of products, which is not described explicitly.

\subsection{Money vs.\ wealth}  \label{Sec:Wealth}

As explained above, money and physical products belong to different layers of the economy.  However, they are often combined into a single variable $w$ called wealth.  The wealth $w_i$ of an agent $i$ is defined as $w_i=m_i+\sum_\alpha p_\alpha v_i^{(\alpha)}$, where $p_\alpha$ is the monetary price per unit volume (per item) of the product $\alpha$, and $v_i^{(\alpha)}$ is the volume (the number of items) of this product owned by the agent $i$.  The prices $p_\alpha$ are conversion coefficients from physical to monetary units.

Because the concept of wealth combines two fundamentally different layers, not surprisingly there are various problems with this concept.  Let us consider wealth change in the example of investment in gold.  The total initial wealth in the system, \$400, is the sum of the total money \$300 owned by the three agents and the monetary value of gold, \$100, as observed in the transaction between agents 1 and 2.  However, the final wealth, \$500, is greater, because the higher price of \$200 was observed in the transaction between agents 1 and 3.  It is somewhat paradoxical that, while the amounts of money and gold (as measured by weight) have not changed, the total wealth has increased.  In the first transaction between agents 1 and 2, $w_1$ and $w_2$ did not change, because each of them exchanged money/gold for the equivalent amount of gold/money worth of \$100.  Similarly, $w_1$ and $w_3$ did not change in the second transaction worth of \$200.  Apparently, the wealth $w_1$ of agent 1 changed between the first and second transactions, when the price of gold increases from \$100 to \$200, because the agents changed their opinion how valuable is gold.  In conclusion, wealth does not change in transactions, but changes between transactions, either because of price change in $p_\alpha$ or volume change in $v_i^{(\alpha)}$ due to physical production (or consumption).  Thus, fallacies \#1 and \#2 do not apply to wealth: wealth can change as a result of investment (due to price change) or production (due to volume change).  However, this brings us to the next fallacy.

\subsubsection*{Fallacy \#4: Kinetic models of wealth transfer (as opposed to money transfer)}

Here I would like to criticize my fellow econophysicists.  As explained in Sec.\ \ref{Sec:Personal}, a kinetic gas-like model with collisions/transactions between agents was introduced in \cite{Yakovenko-2000} as a model of money transfers among agents (in exchange for goods and services not specified explicitly).  However, many other econophysics papers (see reviews \cite{Yakovenko-2009,Chakrabarti-2013}) apply similar kinetic models to conservative wealth transfer between agents using the rules $w_i'=w_i-\Delta$ and $w_j'=w_j+\Delta$.  Already the earlier paper \cite{Redner-1998} talked about wealth and assets transfers, but money was not mentioned.  Although the models for transfer of $m$ or $w$ are mathematically the same, there is important conceptual difference between wealth and money.  As explained above, the wealth $w_i$ of an agent $i$ does not change in a transaction, because money is exchanged for goods of equivalent value, which means that $\Delta=0$ in all transactions.  On the other hand, wealth does change between transactions due to price changes or production, which are not included in these models, so the total wealth is actually not conserved, but these models are conservative.  In my opinion, the terminology in these models should be changed from wealth to money.  Otherwise, the word ``wealth'' makes them conceptually confusing and misleading.  Some papers argue that wealth transfer between two agents happens due to deceit and stupidity (resulting in wrong price paid) or robbery and theft (forceful or concealed seizure of property).  Although such incidents do happen, they play a minor role relative to regular market transactions in an orderly society.

\subsection{Exogenous money creation by the state}  \label{Sec:State}

As discussed in Sec.~\ref{Sec:Nature} and argued by MMT \cite{MMT}, a sovereign state has the authority to issue fiat money from scratch.  Obviously, such emission (``printing'') of new money by the state changes the total amount of money $M$ in circulation, so it is a mechanism for non-conservation.  However, the terminology needs further refinement here.  When discussing conservation laws in physics, it is necessary to draw a boundary to define a system.  Then a conservation law, e.g.\ of energy, says that the total energy inside the system cannot change due to any processes inside the system and can only change due to energy transfer across the boundary.  For example, if we choose the Earth as the system, its total energy can only increase due to energy flux from the Sun and decrease due to radiation into space.  Although the energy of the Earth may change if the two fluxes are not equal, this example is completely consistent with the conservation law in physics.  Similarly, we can define the regular economic system as consisting of all private agents of the economy, whereas the state is outside the system.  As in the Earth example and following MMT \cite{MMT}, ``horizontal'' transactions among the agents within the system do not change total money, but, in ``vertical'' transactions between the system and the state, new money flows across the boundary into the system.

Why would the state want or need to inject new money into the system?  One reason, out of many, is population growth.  If population $N$ grows while money $M$ is constant, the money per capita (the money temperature) $T_m=M/N$ would decrease.  Since money becomes scarcer on per capita basis, its buying power would increase, so prices would decrease, which is called deflation.  Once the agents realize that prices are decreasing, they would tend to hoard money, because they can buy more products for the same money in the future.  Hoarding would decrease the amount of money in circulation, which would further exacerbate deflation and may suffocate the economy.  So, arguably, the state should increase $M$, at least, in proportion to $N$ to prevent $T_m$ from decreasing.  The monetary school of economics \cite{McConnell} proposed a monetary rule of steady money injection on a regular schedule, arguing that a moderate inflation would stimulate spending and prevent money hoarding.

One may also argue that the new, younger generation is ``entitled'' to the same amount of money per capita as the previous, older generations.  However, it is not a good idea just to give lump sums of money to young people.  It is better to follow the principle that money must be earned by work.  So, the best way for the state to inject new money is by funding public infrastructure projects for the benefit of the whole society.  These projects can be also paid by taxes collected from the population.  But a proportion of taxes and new money for public infrastructure funding is a technical and practical matter, not a dogma like the ``balanced government budget''.

The prime example of public infrastructure is military spending.  It is hard to assess consumer value of the enormously expensive intercontinental ballistic missiles, which nobody ever wants to use.  Only the government can pay for them, and the military-industrial complex has been the prime driver of the American economy for many years, creating various high technologies along the way.  Arguably, the greatest challenge for the humankind in the 21st century is an orderly transition from fossil fuels to renewable energy, so the state should actively fund such a transition.  Given that trillions of dollars of new money have been created in recent memory to bail out financial institutions and to pay for wars, the laments that ``there is not enough money in the budgets to pay for renewable energy'' or ``renewable energy is not affordable because it is too expensive'' sound quite ridiculous. 

Within the state, the government, which is the executive branch, is usually separated from the central bank, which has monetary authority.  Thus, the government cannot issue new money on its own and can only raise money for its budget by collecting taxes and fees or by borrowing, e.g.\ by issuing treasury bonds.  Private agents would give their money to the government in exchange for treasury bonds, and the government would use this money to pay for public projects.  Thus, government borrowing from the agents does not change the total money in the system.  At the bond maturity date, a reverse process is supposed to happen.  The government would need to collect enough taxes and fees to pay money back to bond holders.  However, the government budget in USA is almost always in deficit, so the Department of the Treasury keeps issuing new bonds to pay back the earlier ones.  New money is injected into the system when the Federal Reserve (the central bank) buys the treasury bonds.  The whole process looks like a charade.  The Federal Reserve is not permitted to buy bonds directly from the Treasury and can only buy then on the open market from commercial banks, thus producing the same net outcome, but paying handsome fees to the banks for ``intermediation''.  Moreover, a footnote in \cite[p.\ 267]{McConnell} explains that the Treasury pays interest on the bonds to the Fed; the Fed records these payments as ``profit'' and then, being a non-profit public agency, returns them to the Treasury, which records them as ``revenue''!  The net result is that the treasury bonds held by the Fed represent an interest-free loan to the Treasury that never has to be paid back.  Thus, the debt from the Treasury to the Fed should be subtracted and erased from the net government debt \cite{Duncan}.  The recent massive purchase of the treasury bonds by the Fed is monetization of the government debt and injection of new money into the system.  It is very unlikely that the Fed will ever sell these bonds, thus removing money from the system.  Moreover, it is likely that the Fed will need to buy further massive amounts of treasury bonds from the Social Security Administration (SSA) to enable retirement payments for the baby boomer demographic wave.  The SSA will need a bailout by the Fed, because it spent all revenue from social security taxes to buy treasury bonds, thus funding operating budget of the government.  Somebody has to pay cash for these bonds, most likely the Fed, because the government in perpetual deficit cannot do that.  Arguably, one reason for the prolonged recession in Europe is resistance by the European Central Bank to monetize government debt.

In a typical manner of checks and balances in USA, new money can be injected only upon action and consent of three state bodies.  The executing branch (the government) has to propose a deficit budget and request bond issuance, the legislative branch (the Congress) has to approve this request, and the Fed has to buy these bonds.  Lately, the three bodies often acted without coordination and in opposite directions.  The Congress pressed for austerity, demanding budget cuts and refusing to raise debt ceiling for the Treasury, whereas the Fed pursued easy money policy and bought a lot of treasury bonds.  Due to the massive injection of new money, one would expect to see significant inflation, for which there are many examples in countries around the world.  Yet the consumer price index in USA has not changed much.  One possible explanation is globalization of the world economy \cite{Duncan}.  As a legacy of the Bretton Woods agreements, the American dollar is the world reserve currency, which is used formally or informally in many countries.  So, the new money injected by the Fed do not necessarily stay within USA, but spread around the world by the globalized financial system.  Thus, it would be more appropriate to consider the huge global population as the number of agents $N$, rather than the population of USA.  In the past, when national economies were more closed, emission of money by a national central bank was likely to cause inflation.  But in the modern globalized world, emission of money by the Federal Reserve stimulates a Turkish economic bubble \cite{Colombo} more than the American economy.

\subsection{Debt in peer-to-peer lending}  \label{Sec:Peer}

A widely held opinion is that debt has something to do with money creation.  So, we examine the question of debt in the rest of the paper.  The ``rich'' agents who accumulated high money balances, well above their customary level of consumption, may not know what to do with this money, but would not want to just give it away.  Instead, they may decide to lend this money temporarily to other, ``poor'' agents, who have low or zero money balances.  First, let us examine such peer-to-peer lending without banks.  

\subsubsection*{Fallacy \#5: Money grows as a result of peer-to-peer lending}

Agent 1 would give money $\Delta$ to agent 2 in exchange for a promissory note (an IOU), where agent 2 promises to return the money $\Delta$ back to agent 1 at a specified time in the future, with an extra monetary interest $\delta\Delta$.  Nominally, this transaction looks similar to what was considered at the beginning of the paper.  The money balances of the agents change as $m_1'=m_1-\Delta$ and $m_2'=m_2+\Delta$ in exchange for the financial asset (the IOU), rather than physical goods.  The total money is conserved, because agent 1 does not have money $\Delta$ while agent 2 has it, and the IOU does not circulate, because it is a personal agreement between two agents.

\subsubsection*{Fallacy \#6: Money grows as a result of interest}

In order to return the principal $\Delta$ with the interest $\delta\Delta$ to agent 1, agent 2 has to obtain this money from other agents.  As in the example of investment in gold, the balance of agent 1 would increase to $m_1''=m_1+\delta\Delta$ after repayment of debt with interest, but the increase $\delta\Delta$ has to come from other agents in the system.  So, interest on debt does not increase total money and does not violate conservation law of money.

\subsubsection*{Fallacy \#7: Money is debt}

Superficially, the act of borrowing looks somewhat similar to the positive-negative pair creation in LETS discussed in Sec.\ \ref{Sec:Nature}.  Agent 1 delivered goods to agent 2, and their balances became $\tilde m_1'=+\Delta$ and $\tilde m_2'=-\Delta$.  The negative balance of agent 2 looks like debt, because it implies a promise to deliver equivalent goods to agent 1 sometimes in the future, with cancellation of the balances upon delivery.  However, money and debt are not the same thing, because they have important differences:
\begin{itemize}

\item Debt creates a ``string'' (or a ``chain'') connecting the two agents until the debt is repaid.  In contrast, a money transaction is final, and the agents have no further obligations or connections, once goods are delivered and money paid.

\item Debt is personal, whereas money is anonymous.  The names of agents 1 and 2 are written in the IOU agreement between them, but not on dollar bills.

\item Debt has a due date for repayment, whereas money has no time stamp.

\item There are legal penalties for not repaying debt, but no such things for money.

\item Debt usually carries interest, but there is no such thing for money.

\end{itemize}
In short, debt is a promise to pay money, but not money itself.  For example, suppose the agents use bitcoins as their monetary tokens.  They can still create a system of lending and borrowing on top of their monetary system.  An agent 1 would transfer bitcoin to agent 2 in exchange for an IOU, where agent 2 promises to return bitcoin to agent 1 at a specified time, possibly with extra bitcoin interest.  Thus, lending and borrowing can be thought of as the third, promissory, layer of the economy, on top of the monetary level, where the actual money moves.

\subsection{Debt as particle-antiparticle pair creation}  \label{Sec:Pair}

Let us consider debt from the point of view of financial wealth (net worth) $\tilde w$.  Let us define the latter as the sum $\tilde w_i=m_i+d_i$ of money and money obligations, i.e.\ debt.  Here $d_i=\sum_j d_{ij}$, where $d_{ij}$ represents the debt owed by agent $j$ to agent $i$, and the matrix is antisymmetric: $d_{ji}=-d_{ij}$.  Positive $d_i>0$ means that other agents owe debt to agent $i$, and negative $d_i<0$ means that agent $i$ owes debt to them.  Suppose agent 1 has zero initial money balance $m_1=\tilde w_1=0$ and then borrows the amount $\Delta$ from agent 2, whose initial money balance is $m_2=\tilde w_2$.  After the transaction, agent 1 has money $m_1'=+\Delta$ and debt obligation $d_1'=-\Delta$, but the net worth is still zero $\tilde w_1'=+\Delta-\Delta=0$.  Net worth of agent 2 also remains the same: $\tilde w_2'=m_2'+d_2'=m_2-\Delta+\Delta=\tilde w_2$.  Thus, borrowing looks similar to particle-antiparticle (e.g.\ electron-positron) pair creation in physics with net zero electric change.  The act of borrowing does not change net worth of an agent, but allows it to become negative by removing the boundary condition.  Indeed, after agent 1 spends some of the money $m_1'=+\Delta$ to buy goods and services from other agents, the balance becomes negative $w_1''<0$, because the debt obligation $d_1'=-\Delta$ stays with the agent.

\subsubsection*{Fallacy \#8: Debt stabilizes by itself}

Suppose the agent-based simulation \cite{animation} described at the end of Sec.\ \ref{Sec:Nature} is repeated for net worth $\tilde w$ instead of money $m$, starting from the equal money balances and no debt.  Now, when the agents need to buy goods or services, but do not have enough money, they can borrow money from other agents by issuing debt obligations.  As a limiting case, suppose there is no restriction on debt.  Then the initial delta-function distribution $P_0(\tilde w)=\delta_{\tilde w,\langle m\rangle}$ would keep broadening in a Gaussian manner and would never stabilize, because negative balances are now permitted, and the boundary condition at $\tilde w=0$ is removed.  Some agents would have increasingly negative balances due to accumulation of debt, and other agents have increasingly positive balances by counting the debt owed to them as an asset, while the total net worth in the system $\tilde W$ remains constant.  The entropy $S$ of the distribution $P(\tilde w)$ would increase indefinitely without saturation.  The increase of entropy indicates that the growth of debt is an irreversible process, similar to diffusion.  Once debt is created, it cannot be paid back for the whole ensemble, although each individual agent has a non-zero probability to pay back the debt.  In conclusion, debt does not stabilize by itself, unless a debt restriction is imposed, and there is no such thing as equilibrium debt. 

\subsubsection*{Fallacy \#9: Interest stabilizes debt}

If interest is introduced in the above consideration, it would further destabilize the system.  Negative balances would become more negative, and positive balances more positive, so the positive and negative parts of the probability distribution $P(\tilde w)$ would diverge (rather than converge) even faster.  So, interest accrual, by itself, does not stabilize the system, but destabilizes it.  Only a debt restriction can stabilize it.

\subsubsection*{Fallacy \#10: Economic models where debt is always repaid as promised}

If debt were always paid back as promised, it would have not mattered much and could have been ``integrated out'' and omitted from consideration, as is often done implicitly in many economic models.  But, as explained above, some agents will not be able to pay it back, because debt creation is a statistically irreversible process.  An important question is what happens when the due date comes, but an agent cannot pay back the debt.  One possibility is that the agent would declare bankruptcy and have the debt erased by a judge.  However, another agent counted this debt obligation as an asset, so it will be also erased.  This process is similar to annihilation of particle-antiparticle pairs connected by the strings of debt.  Repayment of debt would also annihilate it.  Another option is that the agent would take new debt to pay the old debt.  Debt repayment would be postponed, and the total debt in the system would continue to grow until it reaches a critical value, then triggering a massive cascade of bankruptcies.  This would be the Hyman Minsky moment \cite{Minsky}.  Physics analogies of this phenomenon include stick-slip motion and earthquakes, where stress gradually builds up until tectonic plates move and release the stress, or forest fires, where forest grows until reaching a critical mass and then is swept by fire.  If borrowing and defaulting happened in a random uncorrelated manner among agents, the system could have achieved a stationary steady state.  However, in reality, debt growth and collapse happens in a highly correlated and synchronized manner, when many agents do the same thing, and a strong collective mode emerges in the system.  A detailed conceptual understanding and mathematical description of switching between regimes of debt expansion and debt contraction, qualitatively described by Minsky \cite{Minsky}, is an open problem for econophysics.  From physics point of view, it represents a spontaneous time-translation symmetry breaking, where the system develops an oscillatory (possibly chaotic) behavior in time, as opposed to a stationary, time-independent equilibrium.  There are attempts to describe this behavior using a system of nonlinear ordinary differential equations \cite{Keen}, but a bigger challenge is to observe emergence of this collective mode in a statistical ensemble of many interacting agents.  

One well-known mechanism for synchronization of debt is inflation of collateral prices.  To feel safer about lending money, a lender may demand that a borrower pledges to transfer a tangible physical object (e.g.\ a house), called collateral, in case of default on the debt.  The maximal loan would be limited by the ``market price'' of the collateral.  But lending money for buying houses affects their market prices, so the price of the collateral would go up, and lending would increase further, and so on to create a credit bubble.  However, Basel regulations of banking issued by the Bank for International Settlements failed to recognize this circular logic in ``mark to market'' accounting.  The problem is similar to the example of investment in gold discussed in Sec.\ \ref{Sec:Wealth}, but with the added run-away instability due to lending.

\subsection{Banks}  \label{Sec:Banks}

Peer-to-peer lending is somewhat similar to barter, in the sense that it involves pairwise promissory agreements of repayment between agents.  At a higher level, lending and borrowing can be aggregated and anonymized by the banks (that is commercial banks, as opposed to the central bank, which is a public non-profit agency of the state).  The agents willing to lend would deposit their money into a bank, and the bank would lend this money to the agents willing to borrow.  This arrangement scales up the operation by aggregation and replaces pairwise debt agreements between the agents by one-to-many debt agreements with the bank.  As a result of the change in the lending network topology, the depositors do not know any more to whom their money is lent, which implies loss of information and increase in informational entropy.

Moreover, in contrast to peer-to-peer lending, the depositors can still use their money, even though it is lent to other agents.  The bank creates checking accounts for the depositors, so they can write checks instead of using the actual money, whereas the deposited money is placed by the commercial bank on reserve in the central bank.  Thus, a commercial bank has two faces, one toward the customers (depositors and borrowers) and another toward the central bank, through which interbank transactions (such as check clearing) are performed.  When agent 1 with an account in bank A writes a check to agent 2 with an account in bank B, the check is cleared via the central bank, and the corresponding amount of reserves is transferred from the account of bank A to the account of bank B in the central bank.  Thus, interbank transactions form yet another layer of the monetary economy on top of the inter-agent transaction layer.  Now the conservation of law of money applies to the bank reserves, which are only transferred, but not created in interbank transactions.  

Very importantly, because of aggregation, banks can support the volume of transaction among the agents much greater than the amount of bank reserves.  If agent 1 writes a check to agent 2 who has an account in the same bank, check clearing does not involve transfer of reserves, and the bank simply changes the digits on the accounts of the agents.  Moreover, when many customers of banks A and B write checks in both directions, these checks partially cancel, so the daily net transfer of reserves can be much smaller than the total volume of transactions.  In addition, banks can temporarily borrow reserves from each other and from the central bank (so, the whole story of peer-to-peer landing described in Sec.\ \ref{Sec:Peer} is repeated at the level of banks, instead of the agents).  

\subsubsection*{Fallacy \#11: Banks create money (actually, they create debt)}

The ability to support a larger volume of transactions with a smaller amount of reserves is called the money multiplier effect \cite{McConnell}.  When agent 1 deposits her money into a bank, she thinks that she still has this money, because she can write checks on this amount.  However, agent 2, who borrowed money from the bank, also thinks he has that money, because he can also write checks on that amount.  It looks like the amount of money in the system of agents has doubled, as described in Ch.~14 ``How Banks Create Money'' in textbook \cite{McConnell}.  However, the more appropriate title for this chapter should have been ``How Banks Create Debt'', because the increase of money in circulation among agents is exactly equal to the increase of debt in the system.  When an agent requests a loan from the bank, the bank creates a checking account for the agent with the corresponding balance, listed as liability on the balance sheet of the bank, and obtains an IOU from the agent, listed as an asset on the balance sheet of the bank.  As in Sec.\ \ref{Sec:Pair}, bank lending is analogous to particle-antiparticle pair creation.  Agent 2 is still connected to the bank by the string or chain of debt, as discussed in Sec.\ \ref{Sec:Peer}.  Borrowing money is different from receiving money as payment, gift, or grant.

Once I had a discussion with the fellow econophysicist Joseph McCauley, who argued, like many economists, that he can create money with a few clicks on a computer keyboard.  Later I invited him to give a talk at the annual conference of the American Physical Society.  He responded that he is currently in Europe, but transatlantic air tickets are very expensive, and he has no funding for travel.  So, I recommended him to do a few clicks on a computer keyboard and create money for his travel to the conference.  The problem here is obvious.  It is true that debt can be easily created with a few clicks, but debt is not the same as money, see Fallacy \#7.

Notice also that debt creation for bank customers does not change reserves of the banks held in the central bank, which are still subject to the conservation law.  So, the ``endogenous'' creation of money/debt by commercial banks for their customers is not the same as the ``exogenous'' money injection by the state, described in Sec.\ \ref{Sec:State}, which has no strings, chains, or debt attached to it.  In MMT \cite{MMT}, these two types of transactions are classified as horizontal and vertical.

\subsubsection*{Fallacy \#12: Reserve requirements limit maximal debt}

How much total debt can the banks create in this manner for the agents?  By regulations, banks must have the required reserves equal to the fraction $R$ of the total deposits, called the required reserve ratio.  If the original money $M_0$ (the money base) is deposited into banks, the amount available to lending is equal to $(1-R)M_0$, because $RM_0$ must be held in reserve.  The borrowed money $(1-R)M_0$ would be deposited into banks again, and the amount available for further lending is $(1-R)^2M_0$.  Continuing this geometrical progression, we find that the money multiplier for maximal permitted lending is $M/M_0=1/R$ \cite{McConnell}, which implies the debt increase by the factor of $(1-R)/R$ \cite{Yakovenko-2009}.  In addition to required reserves, banks also must have some excess reserves to clear interbank transactions.

However, in reality, both limiting factors for the debt are actually soft, not hard, constraints.  The reserve requirements apply only to some financial institutions (primarily retail banks insured by the Federal Deposit Insurance Corporation, FDIC), but not to investment banks, hedge funds, shadow banks, etc., and do not apply to certain types of accounts, e.g.\ business savings and time deposits \cite{McConnell}.  Some countries have no reserve requirements at all \cite{Reserve}.  So, reserve requirements do not impose an absolute limit on maximal debt created by the banks, which is more of a myth perpetuated by undergraduate economics textbooks \cite{McConnell}.  Banks have even more flexibility with excess reserves and can find various ways to minimize excess reserves and maximize lending, if they want to.  For example, consolidation and mergers of the banks resulted in fewer banks and more customers having accounts in the same bank.  Check clearing between these customers does not require excess reserves for transfers.

\subsection{Problems with run-away debt}  \label{Sec:Run-away}

Fallacy \#12 indicates that banks can create unlimited debt.  However, as discussed in Sec.\ \ref{Sec:Nature}, a limit on negative balances is essential for stability of the system.  If agents can borrow unlimitedly, they can simply pay for goods and services with borrowed money and not contribute in return, which would be free riding.  Moreover, if the banks can ``create money'' as easily as the textbook describes \cite{McConnell}, they would also create money for themselves, not only for their clients.  Even worse, they may actually use their clients as pawns and smoke screen to conceal and obfuscate money creation for themselves.  It is not just a fantasy.  The textbook \cite[p.~271]{McConnell} says that federal investigators found criminal conduct in 40\% of the failed Savings and Loans (S\&L) institutions in the late 1980s and early 1990s.  Financial institutions ``learned'' from their previous mistakes and made sure that no criminal investigations were conducted at all in the much bigger crisis of 2008.  One cannot escape suspicion that the ultimate purpose of the enormously complicated, multilayered ``financial innovations'', such as options, derivatives, default swaps, and various chopped, diced, sliced, and repackaged ``securities'', is to obfuscate and erase information about who is responsible for what.  Creating money for themselves out of nothing is the ultimate free riding, which is the underlying reason for deep hostility that Main Street feels towards Wall Street.

However, although the loans created by the banks may produce an illusion of money bonanza for a while, they still must to be paid back at the due date, at least in theory.  This leads to the Minsky moment of market crash described in Sec.\ \ref{Sec:Pair}.  But now the scale of this crisis is much greater than it would have been for peer-to-peer lending, because of aggregation, consolidation, leveraging, and multiplication by the banks.  Such a crisis (as in 2008) threatens to take down the whole economy, so the central bank has no choice but to step in and save the financial system by absorbing the losses.  This game is played in two steps.  In the first step, the financial system issues huge amounts of loans to borrowers who have no ability to pay back.  Typical examples are subprime mortgages in USA and the Greek debt.  Banks record these loans as assets on their balance sheets.  In the second step, some years later when payments on loans are due, it becomes clear that the borrowers cannot pay.  This creates a huge crisis, so a public monetary authority has to step in.  In USA, the Federal Reserve buys a massive amount of toxic assets from the banks, adding several trillions of freshly ``printed'' dollars to bank reserves (see Sec.\ \ref{Sec:State}).  Similarly, the International Monetary Fund (IMF) gives a new loan to Greece, so that it pays back the debt to private commercial lenders.  In both cases, the net result is that the banks have exchanged the toxic debt assets created in the first step for the real, exogenously created, money from the central monetary authorities in the second step, i.e.\ the second step monetizes the debt created in the first step.  At the end, Greece is nominally in debt to IMF, but still cannot pay, so, most likely, this debt will be eventually written off.  The motto of this game is ``privatize gains and socialize losses''.  For the second step, the bailout by the central bank, to happen, it is essential that the crisis created in the first step has a catastrophic scale, because otherwise the central bank would not bother to step in and would not supply new money.  Thus, one comes to conclusion that creation of large-scale crises is essential part of the current financial system \emph{modus operandi}.  It is needed to transform the endogenously created money/debt by commercial banks in the first step into the exogenously created money by the central bank in the second step.  

The crisis of 2008 is by no means unique.  The S\&L crisis in the late 1980s -- early 1990s required a bailout by the government to the tune of \$200 billion \cite[p.\ 271]{McConnell}.  The discussion of the S\&L crisis in the 20-years-old textbook edition of 1996 \cite{McConnell} is eerily similar to what we heard later about the 2008 crisis.

\subsection{The effect of Quantitative Easing on income inequality}  \label{Sec:QE}

\begin{figure}
\includegraphics[width=\linewidth]{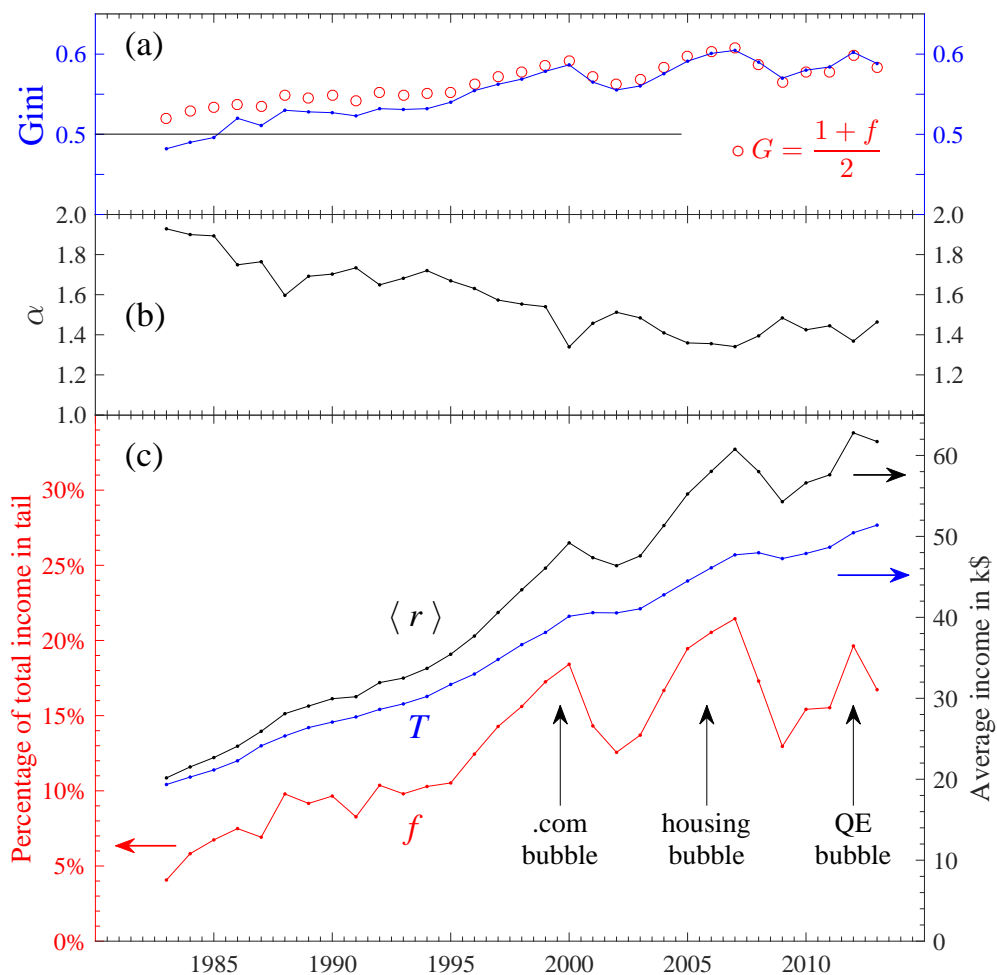}
\caption{(a) The Gini coefficient $G$ for income distribution in the USA in 1983--2013 (connected line), compared with the theoretical formula $G=(1+f)/2$ (open circles).  (b) The exponent $\alpha$ of the power-law tail of income distribution.  (c) The average income $\langle r\rangle$ in the whole system, the average income $T$ in the lower class (the temperature of the exponential part), and the fraction of income $f=1-T/\langle r\rangle$ going to the upper class.}
\label{Fig:1983-2013}
\end{figure}

The described above lend-crash-bailout cycle is a far cry from the original purpose of money, described in Sec.\ \ref{Sec:Nature} as the accounting tool needed to prevent free riding and ensure fair exchange of goods and services.  As described in Sec.\ \ref{Sec:State}, a central bank can print any amount of money, but the important question is to whom this money will be given.  Unlike executive government, which can spend money on roads and bridges and has many other useful channels for injecting new money into the economy, the central bank can only deal with commercial banks.  So, the new money from the Fed inevitably goes to the banks, i.e.\ to the bankers, i.e.\ to the upper class described at the end of Sec.\ \ref{Sec:Personal}, thus increasing inequality in the society. 

This conclusion is supported by the data.  Figure \ref{Fig:1983-2013} shows a two-class analysis of income distribution data from IRS for USA in 1983--2013.  In panel (c), $\langle r\rangle$ is the mean income per capita in the whole system, $T$ is the income temperature in the exponential distribution $P(r)\propto e^{-r/T}$ for the lower class, and $f=1-T/\langle r\rangle$ is the fraction of income going to the upper class.  Panel (b) shows the exponent $\alpha$ of the power-law distribution $P(r)\propto 1/r^{1+\alpha}$ for the upper class.  Panel (a) shows the Gini coefficient $G$ compared with the theoretically derived formula $G=(1+f)/2$ \cite{Yakovenko-2009,Yakovenko-2005a,Banerjee-2010}.  The spikes on the red curve for $f$ in panel (c) indicate sharp increases of inequality due to enhanced share of income going to the upper class.  The 1st spike coincides with the .com bubble in stock market, the 2nd spike with the housing bubble, and the 3rd spike with Quantitative Easing (QE) pursued by the Federal Reserve.  As we see, inequality peaks during speculative bubbles in the markets.  The 3rd spike gives direct evidence that the bailout of the financial system by the Fed resulted in increase of inequality.  This issue has been also brought up in the media \cite{WP-Fed-inequality}.

\subsection{The ultimate source of capitalist monetary profit is state money}  \label{Sec:Profit}

The lend-crash-bailout cycle also gives an insight into the two related questions, which puzzle many people who try to think logically and critically about economic matters.  First, everybody expects their money to grow due to interest.  Where this extra money would come from to pay interest to everybody, given that money is conserved in transactions?  The second question is about capitalist profits, which was framed by Karl Marx in \textit{Capital} as the money-commodity-money, $M-C-M'$, circuit.  Suppose capitalists have the initial stock of money $M$, but would like to have even more money.  (Apparently, pursuit of an ever increasing abstract number is the goal in itself for many people.  The Wall Street people say that the purpose of money is just to keep the score, as in sports competition.)  So, the capitalists spend the money $M$ to build factories, hire workers, buy raw materials, and eventually produce consumer commodity $C$, say cars.  Now they expect to sell the commodity $C$ on the market and end up with a higher money balance $M'>M$, i.e.\ with a profit.  However, to whom would this commodity be sold?  Assuming that it is a consumer product for the masses, like cars, it would be sold to the workers, who can pay for it only as much as they were paid by the capitalists.  So, in aggregate, the net money balance $M'$ of the capitalists after selling the cars cannot exceed their initial balance $M$.  Individual capitalists may achieve profits by creating losses for less successful competitors, but the whole class of capitalists cannot have net profit.

Both interest and profit can be funded, if consumers go into debt and temporarily generate extra money in the system via particle-antiparticle pair creation.  But sooner or later this debt has to be paid back via annihilation process.  The financial system would use various tricks to postpone the day of reckoning.  However, in the same way as prevention of small forest fires results in accumulation of fuel and eventual huge, catastrophic forest fire \cite{GregIp}, the accumulated postponed debt eventually leads to a catastrophic crisis.  At this point, the state (the government and the central bank) have to step in and inject new exogenous money into the system, which ends up producing profit and interest.  So, one comes to conclusion that the ultimate source of capitalist monetary profits is state money.  This conclusion is supported by MMT \cite{Wray}.  However, tellingly, neither MMT nor Minsky are ever mentioned in the 1996 edition of the economics textbook \cite{McConnell}.

\section{Conclusions}  \label{Sec:Conclusions}

Physicists approach economics from a different perspective and with different emphasis, as described in Sec.\ \ref{Sec:Personal}.  Conservation laws play very important role in physics, but are rarely formulated explicitly in economics, even when they actually exist.  My personal response to the title of this special issue ``Can Economics Be a Physical Science?'' is that economics would benefit from closer attention to conservation laws, as physical sciences do.  I have been obsessed with conservation of money for many years.  I have also heard all kind of objections against such a conservation law.  This paper is an attempt to catalog and respond to these objections, as well as to clarify related fallacies and confusion.  Much of this confusion is caused by imprecise language and thinking, substitution of terminology, failure to consider all involved elements, etc.  It may very well be that the arguments presented in this paper suffer from the same deficiencies.  So, in the spirit of the \emph{Discussion and Debate} issue, this is a discussion paper and an invitation for a debate.  If the readers disagree with presented arguments, they are welcome to present counter-arguments in a healthy debate.  Hopefully, a better understanding of the subject would emerge as a result.

Instead of repeating main points of the paper, I would like to mention some important topics not included here and future directions.  The paper focuses primarily on money as an abstract digital tool and does not discuss much the physical world.  However, there are intriguing connections between money, debt, and dwindling supply of fossil fuels \cite{Gail}.  This is a very interesting and important topic, but it is outside the volume and scope of this paper.

I argued throughout the paper that the purpose of money is to prevent free riding and mostly considered the one-to-one network topology of transactions among agents.  This approach may be appropriate for the economy focused on production of multiple copies of physical objects, e.g.\ cars.  However, in the 21st century, the economy is shifting toward production of digital objects: software, digital content, online services, etc.  Digital objects, unlike physical ones, can be copied and replicated at essentially zero cost.  As a result, network topology of economic transactions between producers and consumers changes drastically.  In one-to-many topology, the same software or app is used by dozens or millions or even billions users around the world.  In many-to-many topology, many developers work on a product used by many users.  What is the fair pricing of a digital product in these cases?  Because digital objects can be copied so easily, it is often very difficult or practically impossible to collect fees for a product from all users.  Much of the software and digital content that we use is obtained for free, without payment.  Here I do not mean pirated products, but many products officially offered for free.  Thus, most of us enjoy free riding with respect to digital products.  Perhaps the old-fashioned concept of money and obsession with free riding are not quite suitable for the digital economy.  On the other hand, there are multiple ongoing attempts to invent digital money, such as bitcoin.  The new digital money do not have to replicate old money and could have new digital features non-existent in the old money.  For example, when scholarly publishing shifted from printing on paper to online delivery (from journals as physical objects to digital objects), many new features became available to the authors and readers, but the business model of publishing was also profoundly affected.  In conclusion, money is not a static, but an evolving tool, and it remains to be seen what future money will be.

\begin{acknowledgement}
I thank Benedict Mondal for preparation of Fig.\ \ref{Fig:1983-2013} and Ilya Kuntsevich of Beverly Investment Group, whose book \cite{Kuntsevich} addresses many related issues, for numerous discussions.
\end{acknowledgement}


\end{document}